# Interface-driven magnetocapacitance in a broad range of materials

Mario Maglione

ICMCB-CNRS, Université de Bordeaux 1, 87, Av Dr Schweitzer 33806 Pessac France

maglione@icmcb-bordeaux.cnrs.fr

**Abstract** Triggered by the revival of multiferroic materials, a lot of effort is presently undergoing as to find a coupling between a capacitance and a magnetic field. We show in this report that interfaces are the right way of increasing such a coupling provided free charges are localized on these two-dimensional defects. Starting from commercial diodes at room temperature and going to grain boundaries in giant permittivity materials and to ferroelectric domain walls, a clear magnetocapacitance is reported which is all the time more than a few percent for a magnetic field of 90kOe. The only tuning parameter for such strong coupling to arise is the dielectric relaxation time which is reached on tuning the operating frequency and the temperature in many different materials.

Following the "revival" of multiferroic materials [1], many research groups are looking for a coupling between a magnetic field and the dielectric permittivity of materials. Not only applications but also the basic understanding is the driving force for such an extended effort. Since the number of materials which display simultaneously ferroelectric polarization and magnetic order at room temperature is rather limited [2,3], one is looking for alternative routes. More specifically, concerning the magnetocapacitance, a direct room temperature coupling of an external magnetic field with microscopic polarisabilities is still to be found. This is why integrated [4] and bulk [5] composites have been designed as to increase the density of interfaces between ferroelectric and ferromagnetic materials. In piezoelectric-based devices, the generation of a ac current at the piezo resonance is readily achieved [6]. Very recently this effect driven by elastic deformation has been found in very standard Multi Layer Ceramic Capacitors [7] which opens a broad range of applications. Nonetheless room temperature magnetocapacitance, i.e. the change of a capacitance by a magnetic field is still an open question. Recently, Catalan has shown that non-linear resistance at interfaces can be the right way to find such an effective coupling [8]. In this report, we apply this concept to several and very different materials. In a first step, we recall that non-linear resistance of diodes is of everyday use to sense magnetic field. In the blocking regime, it is thus very easy to observe room temperature 15% magnetocapacitance in 1cent diodes provided the operating frequency is set in the right range. We will then move to so-called "giant permittivity" materials [9-12] which include charged interfaces [13-15]. Setting the temperature, in the right range a magnetocapacitance is observed in the case of $CaCu_3Ti_4O_{12}$, an archetype of giant permittivity materials. At last, we focus on more mesoscopic interfaces which are charged ferroelectric domains walls in Fe doped $BaTiO_3$. A magnetocapacitance is clearly seen in the temperature and frequency range where the domain walls are relaxing. With this collection of results in many different materials we demonstrate that effective magnetocapacitance is a very general trend of charged interfaces materials.

Before describing the bulk of our experimental result, we should make a point as to the technical observation of magnetocapacitance. Indeed, the application of strong magnetic field onto dielectric cells needs some caution. In very standard equipment, we have observed a temperature change of 0.01K when the magnetic field was raised from 0 to 90kOe. Even being small, such an electronic disturbance may lead to artificial magnetocapacitance all the more in temperature ranges where the samples dielectric parameter change very quickly versus temperature. For example we have probed ferroelectric samples which neither include any magnetic component nor do they have charged interfaces. High purity Potassium Tantalate was one of these which undergoes a strong divergence of its dielectric permittivity at T<100K. Under isothermal conditions we achieved artificial magnetocapacitance of several percent which, using a differentiation process, could be ascribed to temperature fluctuations of about 0.01K. This will thus be taken at the resolution limit in the following: if the observed magnetocapacitance at a given temperature could stem from a 0.01K fluctuation, it will simply be discarded.

We now start with the first example which is again not new but which will fix the experimental conditions for the following. First, care was taken as to keep the diode in its blocking regime. In this case, the equivalent circuit of the diode is a capacitor with a given level of losses. On figure 1, we plotted the capacitance and the dielectric losses versus time at several spot frequencies; the magnetic field run is sketched on figure 1a. In a way which is shifted versus frequency f, a change of capacitance $\Delta C(f)/C(f)=(C(9kOe,f)-C(0Oe,f))/C(0Oe,f)$ which we will call magneto capacitance in the following is observed. The maximum magneto capacitance is about -11% for f=1MHz. The dispersion of $\Delta C(f)/C(f)$ versus frequency is a signature of the dispersive behavior of the diode capacitance. The dielectric losses $tg(\delta)$ also experience more than 10% variation under 90kOe also depending on the operating frequency. Both the magneto capacitance and the dielectric losses variations may be ascribed to the conducting charges in the semi-conductors. Interface charges are dynamically recombining at the p-n interfaces and the external magnetic field which changes the free charges trajectories through the Hall effect alters the dynamical capacitance [16]. The diode magneto capacitance may thus be described as resulting from the interaction of the external magnetic field with the free charges accumulated at the p-n interface.

Next, we turn to materials where interfaces are much less defined than in p-n junctions. In the $CaCu_3Ti_4O_{12}$ ceramics, a balance between the inner grain conductivity and grain boundaries barriers is the source of an effective giant permittivity [13]. The very specific feature of this huge permittivity is that it is frequency and temperature independent at room temperature while it relaxes following a Deye type behavior at low temperatures [9]. In these ceramics, we performed a magnetocapacitance experiment at room temperature without any sign of coupling in the limit of the above mentioned temperature fluctuation. When however, the sample temperature was fixed in the relaxation range of 100K, a magnetocapacitance $\Delta C(f)/C(f)$ is recorded (figure 2a). At all temperatures, a maximum magnetocapacitance is observed whose frequency increases with temperature exactly in the same way as the relaxation frequency. In the same time, the dielectric losses $\tan\delta$ displayed a variation $\Delta\tan\delta(f)/\tan\delta(f)$ evolving as a S-shape on increasing the operating frequency (figure 2b). This is nothing but the Kramers Kronig transform of the $\Delta C(f)/C(f)$ curves of figure 2a. To exclude the possible thermal fluctuation as the source of these variations, we simulated a temperature variation of 0.01K and found $\Delta C(f)/C(f)$ and $\Delta\tan\delta(f)/\tan\delta(f)$ at least 10 times smaller than the one plotted on figure 2. Temperature fluctuations are thus not the origin of the observed magneto-capacitance. We thus conclude that the magneto-capacitance in $CaCu_3Ti_4O_{12}$ ceramics is observable in the temperature and frequency range of their dielectric relaxation. Since this dielectric relaxation was ascribed to grain boundary layer acting as dielectric barrier between the conducting grains [13], we propose that the magneto capacitance originates from a similar process as in the diode. Free charges accumulated at the grain boundary barrier are interacting with the magnetic field. Since the dielectric relaxation is observed when the temperature and frequency are tuned as to probe the dynamical motion of these charges among these interfaces, the magneto capacitance is maximal right at the dielectric relaxation. We note that a similar magneto-capacitance at the relaxation frequency was already reported in $LuFe_2O_4$ another example of effective giant permittivity material [14]. Since this material is also belonging to the broad family of grain-boundary layer dielectrics [15], a similar magneto capacitance as in $CaCu_3Ti_4O_{12}$ is not a surprise.

The last example of interface related magneto-capacitance will be taken within ferroelectric single crystals. Indeed, in these materials, sharp domain walls are separating macro or micro-domains of homogeneous polarization. The dynamics of these domain walls is at the origin of the ferroelectric hysteresis loop fully analogous to the ferromagnetic loops [17]. However, unlike ferromagnetic domain walls, ferroelectric domain walls carry a strong elastic energy which depends on the exact crystalline symmetry of the ferroelectric phase. Because of this energy, the motion of the domain walls can have macroscopic contributions to the overall dielectric permittivity of ferroelectrics. In $BaTiO_3$, joint elastic and dielectric experiments clearly evidenced this dynamical contribution resulting in a relaxation with about 1eV activation energy [18]. Moreover, the pining of this motion by charged point defects was pointed out many times. In our present experiment, we want to take advantage of this interaction between domain wall dynamics and charged defects to induce an artificial magneto capacitance. This is why we investigated Fe doped $BaTiO_3$ single crystals where the charged defects originate from the heterovalent substitution of $Ti^{4+}$ cations by $Fe^{3+}$ impurities associated to charged oxygen vacancies. These crystals have been deeply investigated because of their interesting optical properties [19]. The dynamical contribution of domain walls to the dielectric properties was however little investigated. On figure 3a, the dielectric loss of a $BaTiO_3$ crystal containing 0.75 atomic percent of Fe are plotted versus temperature between 300K and 4K. Starting from the high temperature side, we can see two sharp and frequency independent anomalies at about 270K and 170K signing the tetragonal to orthorhombic and the orthorhombic to rhomboedral phase transition respectively. Next, a broad maximum is evidenced whose temperature of occurrence is shifted from 140K to 220K as the frequency is scanned from 1kHz to 1MHz. This relaxation process has activation energy of about 1eV and it is not affected by the ferroelectric transition occurring at 170K. These both features are in full agreement with previous reports on undoped $BaTiO_3$ ceramics showing that it originates from domain wall motion [17]. The magnetic field influence on this relaxation is shown on figure 3b for a single frequency (f=100kHz) at two similar cooling runs, one for H=0 Oe and the other for H=90kOe (figure 3b). Only in the vicinity of the relaxation maximum does the dielectric loss depend on the magnetic field. The same coupling features are observed in the capacitance for cooling (figure 3c) and heating (figure 3d) runs. One can see that the 15% magneto-capacitance is restricted to the temperature ranges where domain wall relaxation occurs. If the operating frequency is changed, the occurrence of magneto-capacitance shifts following the trends of figure 3a. Away from domain wall relaxation, no magneto-capacitance was observed. Again, the charge localization at interfaces model holds for this magneto dielectric coupling. Indeed, we already stressed that the domain wall relaxation is pinned by charged defects. In the same time, the Fe-related charged defects are increasing the conductivity of these single crystals. We thus have again the two ingredients –interfaces and free charges- leading to a magneto dielectric coupling.

In all the materials that we have investigated, the localization of free charges at interfaces is the common driving source for the magneto-capacitance coupling. The free charges may be of different origin and the interfaces of different shape and nature, whenever free charge localization contributes to the effective dielectric permittivity a magneto capacitance is observed close to the space charge relaxation frequency. We thus conclude that any device including at the same time interfaces and free charges may lead to strong magneto-capacitance coupling. The observation of magneto-capacitance in ferroelectric crystal doped with heterovalent impurities may also lead to novel routes towards the coupling between an external magnetic field and ferroelectric polarization.

## Methods

The samples were inserted in a PPMS Quantum Design set up at the end of a modified holder using 4 coaxial cables linked to a HP4194 impedance analyzer through BNC connectors and cables. The samples were hold freely by 2 soft wires at the center of the superconducting coil to avoid any spurious strain contributions. For fixed temperature runs, the impedance was recorded at given time interval in the frequency range 1kHz-1MHz while the magnetic field was raised at a rate of 200Oe/s from 0 to 90kOe and then decreased back to 0 after a stabilization dwell time at the maximum field. Such a magnetic field run is sketched by arrows on figure 1a. For temperature experiments, the magnetic field was fixed while the sample temperature was swept at a rate of 0.1k/min up to 1K/min and the sample impedance recorded at several spot frequencies.

References

[1] Fiebig, M. Revival of the magnetoelectric effect. *J. Phys. D: Appl. Phys*. **38,** R123–R152 (2005)

[2] Nicola A. Hill Why Are There so Few Magnetic Ferroelectrics? J. Phys. Chem. B **104**, 6694-6709 (2000)

[3] Kimura, T., Goto, T., Shintani, H., Ishizaka, K., Arima, T. & Tokura, Y. Magnetic control of ferroelectric polarization. *Nature* **426**, 55-58 (2003)

[4] Zheng, H., Wang, J., Lofl and, S.E., Ma, Z., Mohaddes-Ardabilli, L., Zhao, T., Salamanca-Riba, L., Shinde, S. R., Ogale, S. B., Bai, F., Vielhand, D., Jia, Y., Schlom, D. G., Wuttig, M., Roytburd, A. &#xFFFD; Ramesh, R. Multiferroic $BaTiO_3$-$CoFe_2O_4$ nanostructures. *Science* **303**, 661-663 (2004)

[5] Shi Z., Nan C. W., Jie Zhang, Cai N., and Li J.-F. Magnetoelectric effect of Pb(Zr,Ti)$O_3$ rod arrays in a (Tb,Dy)Fe2 /epoxy medium *Applied Physics Letters* **87**, 012503 (2005)

[6] Srinivasan, G., De Vreugd, C. P., Laletin, V. M., Paddubnaya, N., Bichurin, M. I., Petrov, V. M., and Filippov, D. A. Resonant magnetoelectric coupling in trilayers of ferromagnetic alloys and piezoelectric lead zirconate titanate: The influence of bias magnetic field *Phys. Rev. B* **71** 184423-1 (2005)

[7] Israel C., Mathur N. D. and Scott J. F.A one-cent room-temperature magnetoelectric sensor *Nature* Materials **7**, 93 - 94 (2008)

[8] Catalan G. Magnetocapacitance without magnetoelectric coupling *Applied Physics Letters* **88**, 102902 (2006)

[9] Ramirez A.P. , Subramanian M.A., Gardel M., Blumberg G., Li D., Vogt T. and Shapiro S.M.Giant dielectric constant response in a copper-titanate *Solid State Com*. **115**,217-220 (2000)

[10] Ikeda N., Ohsumi H., Ohwada K., Ishii K., Inami T., Kakurai K., Murakami Y., Yoshii K., Mori S., Horibe Y. and Kito H. Ferroelectricity from iron valence ordering in the charge-frustrated system $LuFe_2O_4$ *Nature* **436**,1136-1138 (2005)

[11] Wu Junbo, Nan Ce-Wen, Lin Yuanhua, and Deng Yuan Giant Dielectric Permittivity Observed in Li and Ti Doped NiO *Phys.Rev.letters* **89**, 217601 (2002)

[12] Raevski I. P., Prosandeev S. A., Bogatin A. S., Malitskaya M. A. and Jastrabik L. High dielectric permittivity in $AFe_{1/2}B_{1/2}O_3$ non-ferroelectric perovskite ceramics .A=Ba, Sr, Ca; B=Nb, Ta, Sb. *J. Appl. Phys.*, Vol. **93**,4130-4136 (2003)

[13] Sinclair Derek C., Adams Timothy B., Morrison Finlay D., and West Anthony R. $CaCu_3Ti_4O_{12}$: One-step internal barrier layer capacitor *Applied Physics Letters* **80**, 2153-2155 (2002)

[14] Subramanian M. A., He Tao, Chen Jiazhong, Rogado Nyrissa S., Calvarese Thomas G., and Sleight Arthur W.. Giant Room–Temperature Magnetodielectric Response in the Electronic Ferroelectric $LuFe_2O_4$ *Adv. Mater.* **18**, 1737–1739 (2006)

[15] Maglione M.to appear in *Springer Series in Materials Science* edited by G. Liu and V.Vikhnin

[16] Sze S M, Physics of Semiconductor devices, Wiley international (1969)

[17] Lines M. E. and Glass A. M., Principles and Applications of Ferroelectrics and Related Materials Clarendon, Oxford, (1979)

[18] Cheng Bo Lin, Gabbay M., Maglione M. and Fantozzi G. Relaxation motion and possible memory of domain structures in Barium Titanate ceramics studied by mechanical and dielectric loss *Journal of Electroceramics*, **10**, 5–18, (2003)

[19] Godefroy G.and Perrot A. Dielectric relaxation in Fe or Nb doped Barium Titanate single crystals *Ferroelectrics* **54**, 87-90 (1984)

Figure Captions

Figure 1: room temperature capacitance (a) and dielectric losses (b) of a commercial diode in the blocking regime at several frequencies versus time. The slopes and central dwell mark the magnetic field ramping and stabilization respectively. The maximum magneto-capacitance is 15% for H=90kOe at f=1MHz

Figure 2: relative variation under a magnetic field of 90kOe of the capacitance (a) and dielectric losses (b) of a $CaCu_3Ti_4O_{12}$ ceramic in the temperature range of its dielectric relaxation. The maximum of magneto-capacitance which can be up to 15% follows the dielectric relaxation frequency which increases when the temperature is raised. At room temperature T=300K no magneto-capacitance is found.

Figure 3: (a) dielectric losses of a $BaTiO_3$:Fe single crystal versus temperature for several frequencies. The two sharp anomalies at 170K and 270K mark the ferroelectric phase transitions while the diffuse and shifted maximum stems from the relaxation of domain walls. (b) In the temperature range of the domain wall relaxation the dielectric losses are strongly depressed by a magnetic field of 90kOe. On cooling (c) and heating (d), a 15% magneto-capacitance is observed still in the temperature range of domain wall relaxation.

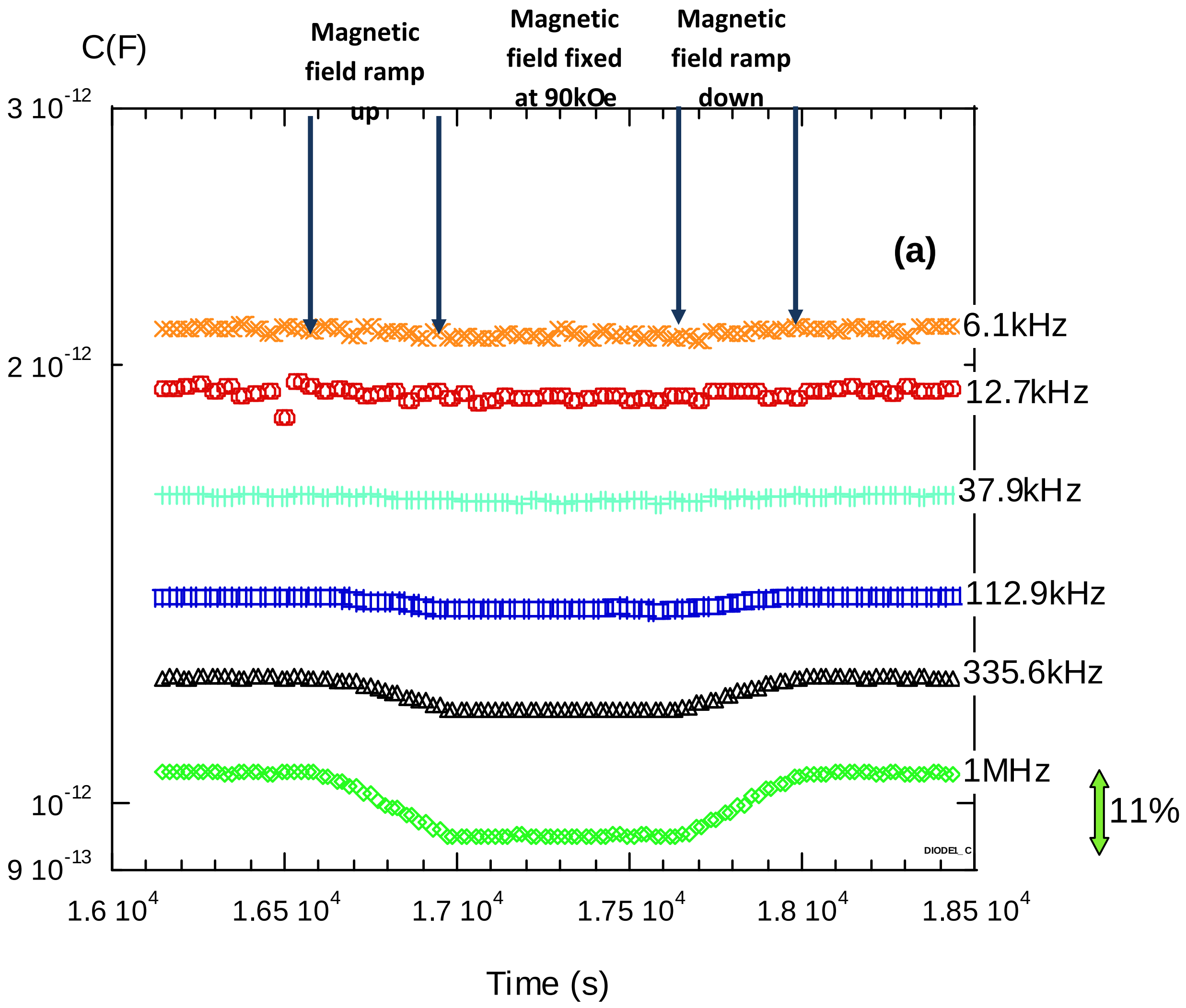

Figure 1a

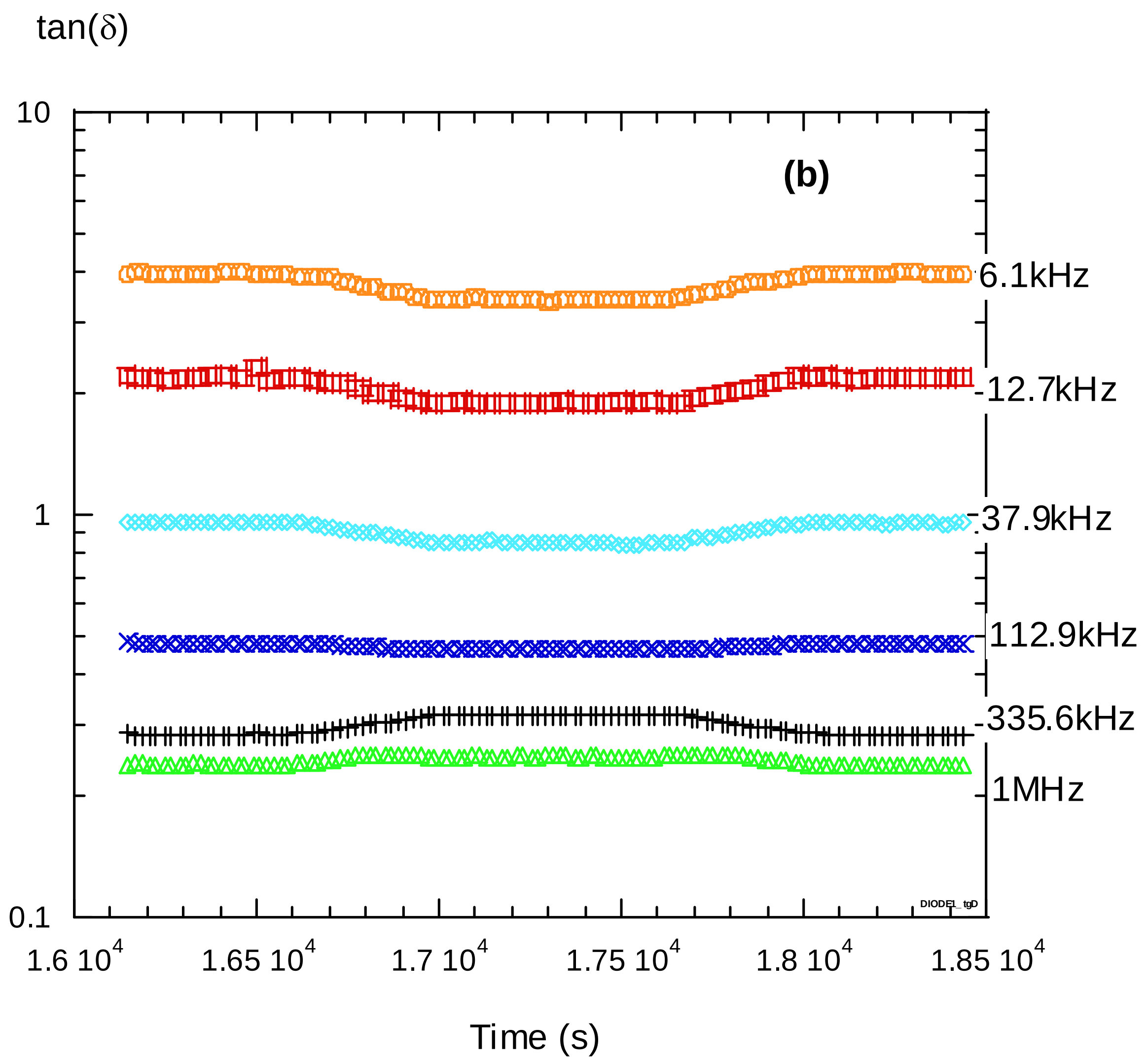

Figure 1b

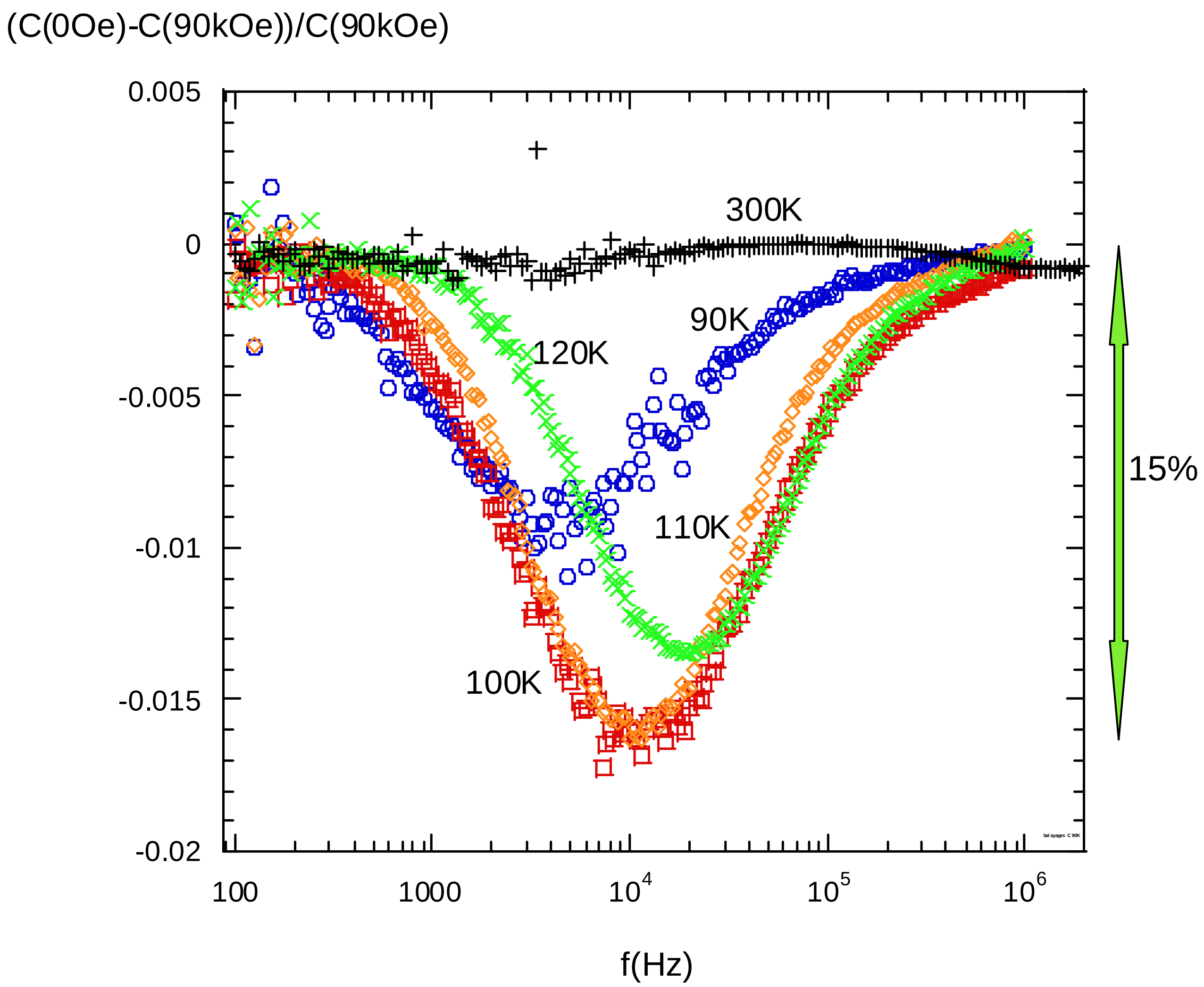

Figure 2a

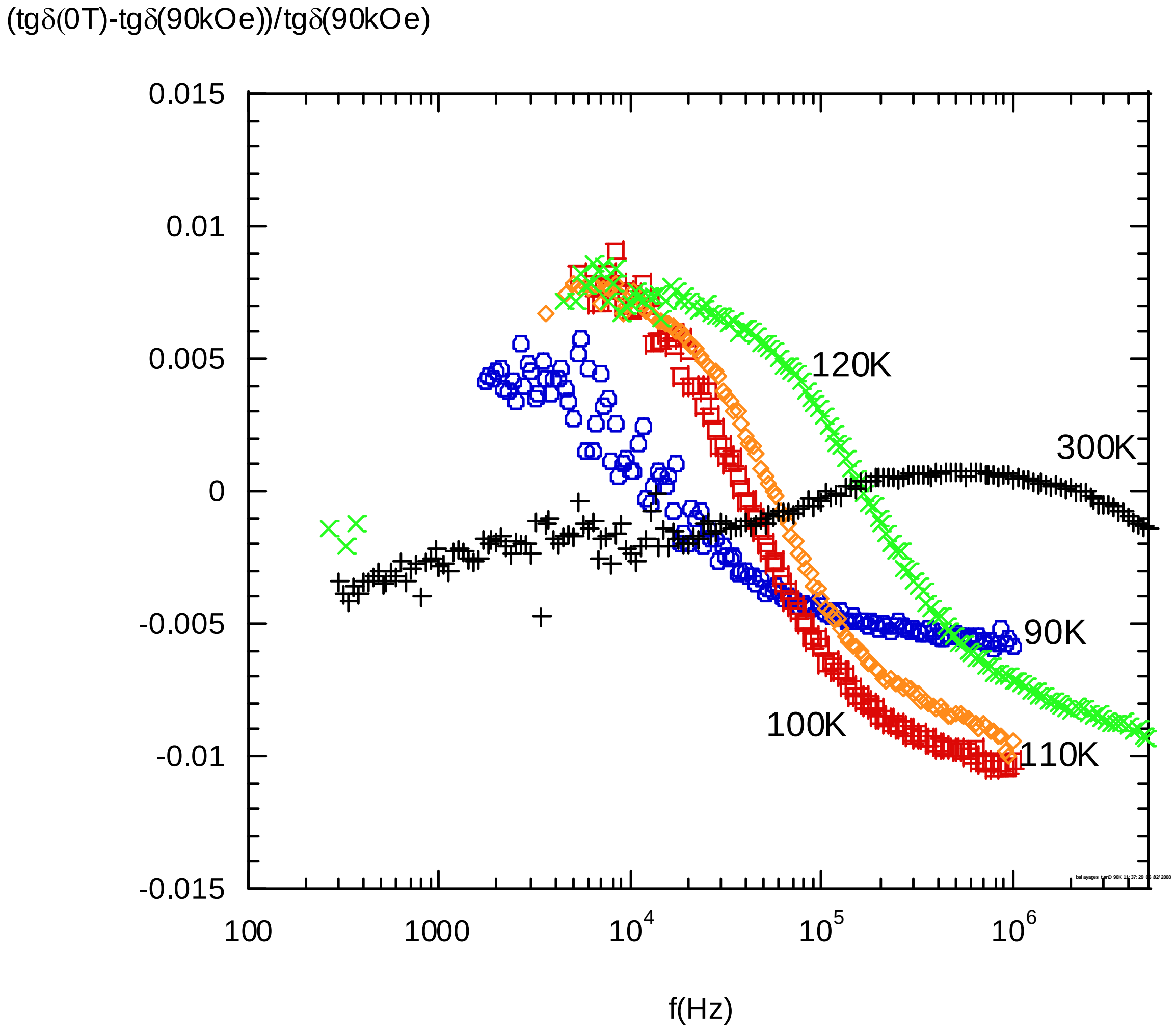

Figure 2b

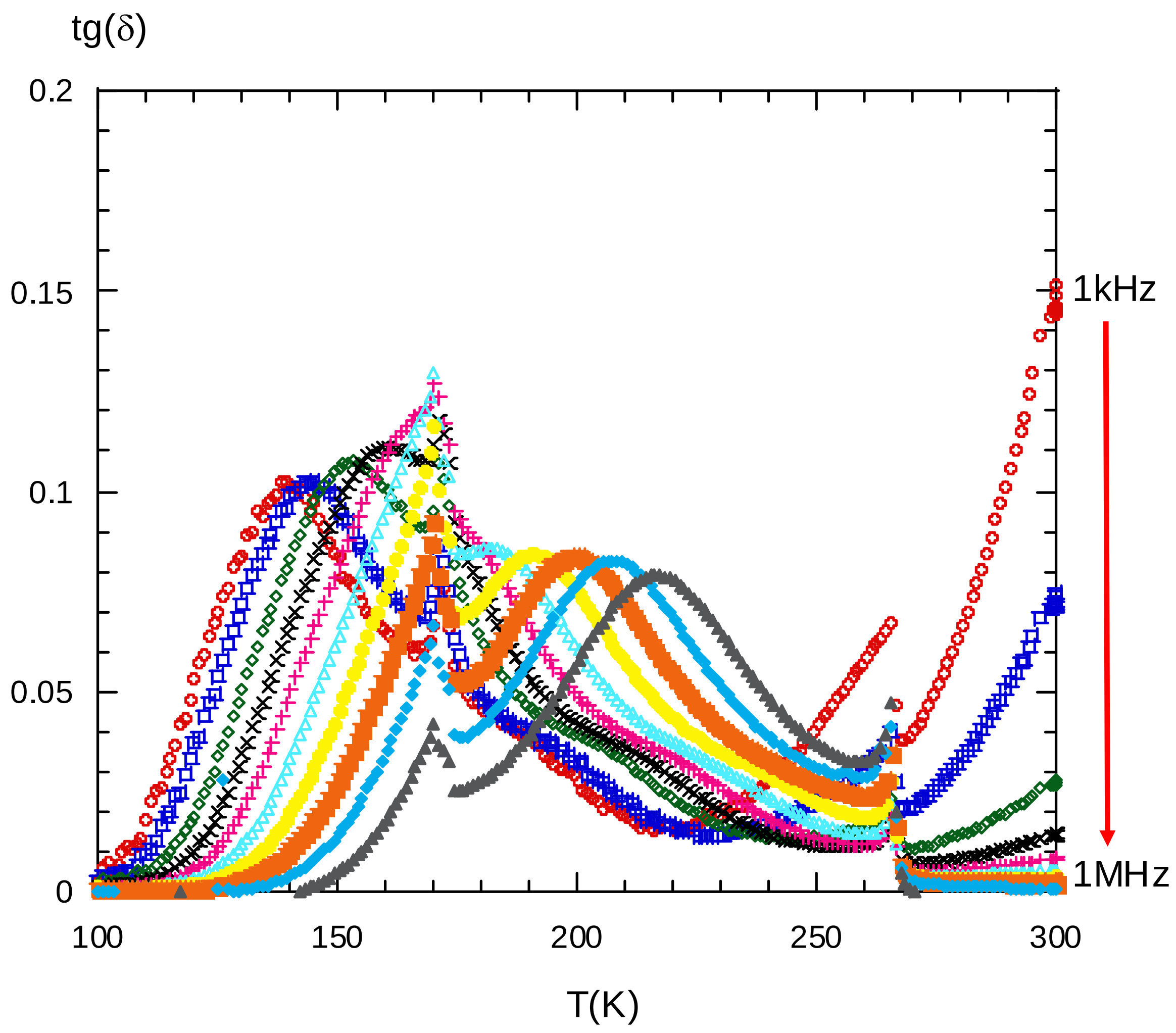

Figure 3a

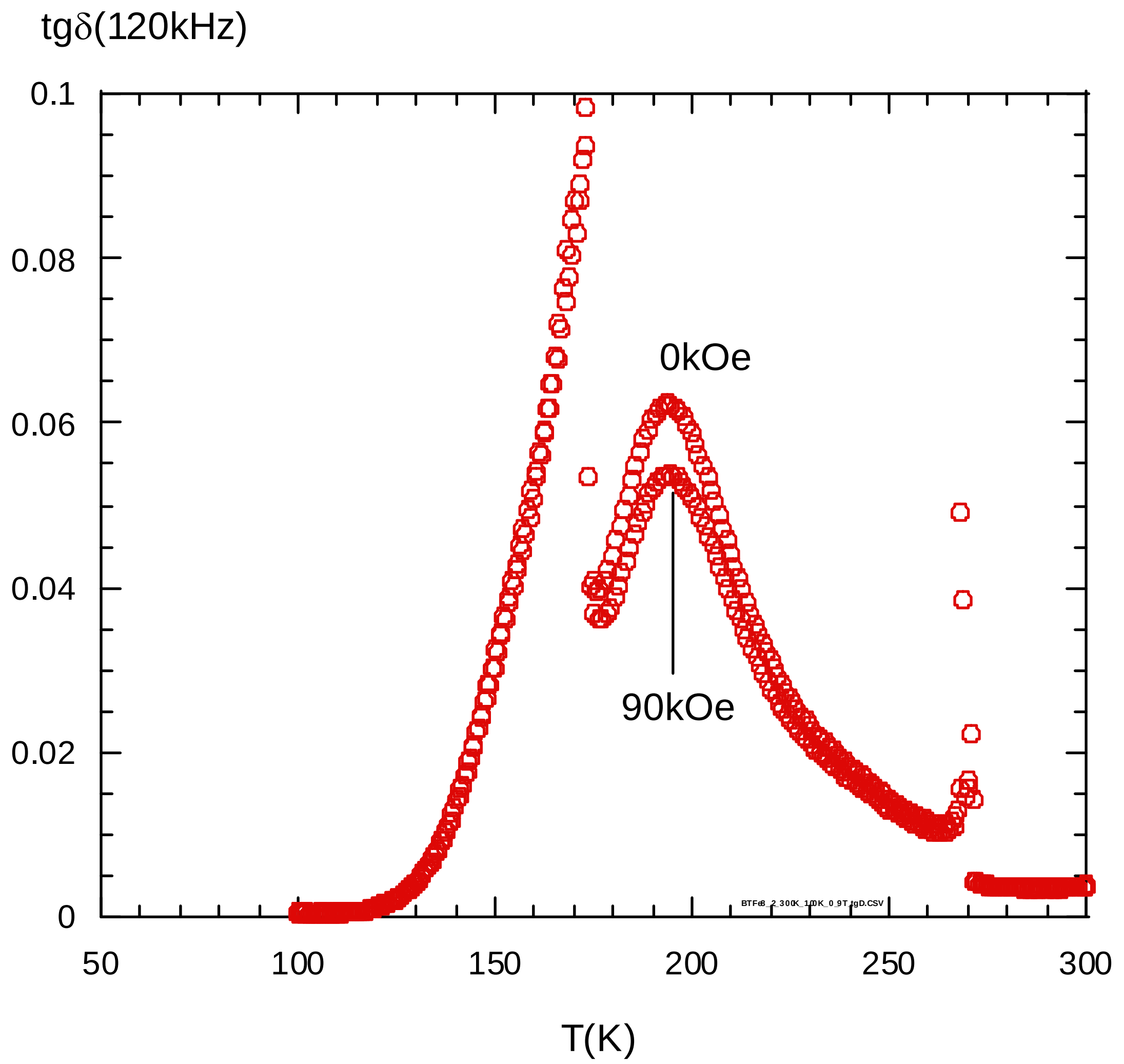

Figure 3b

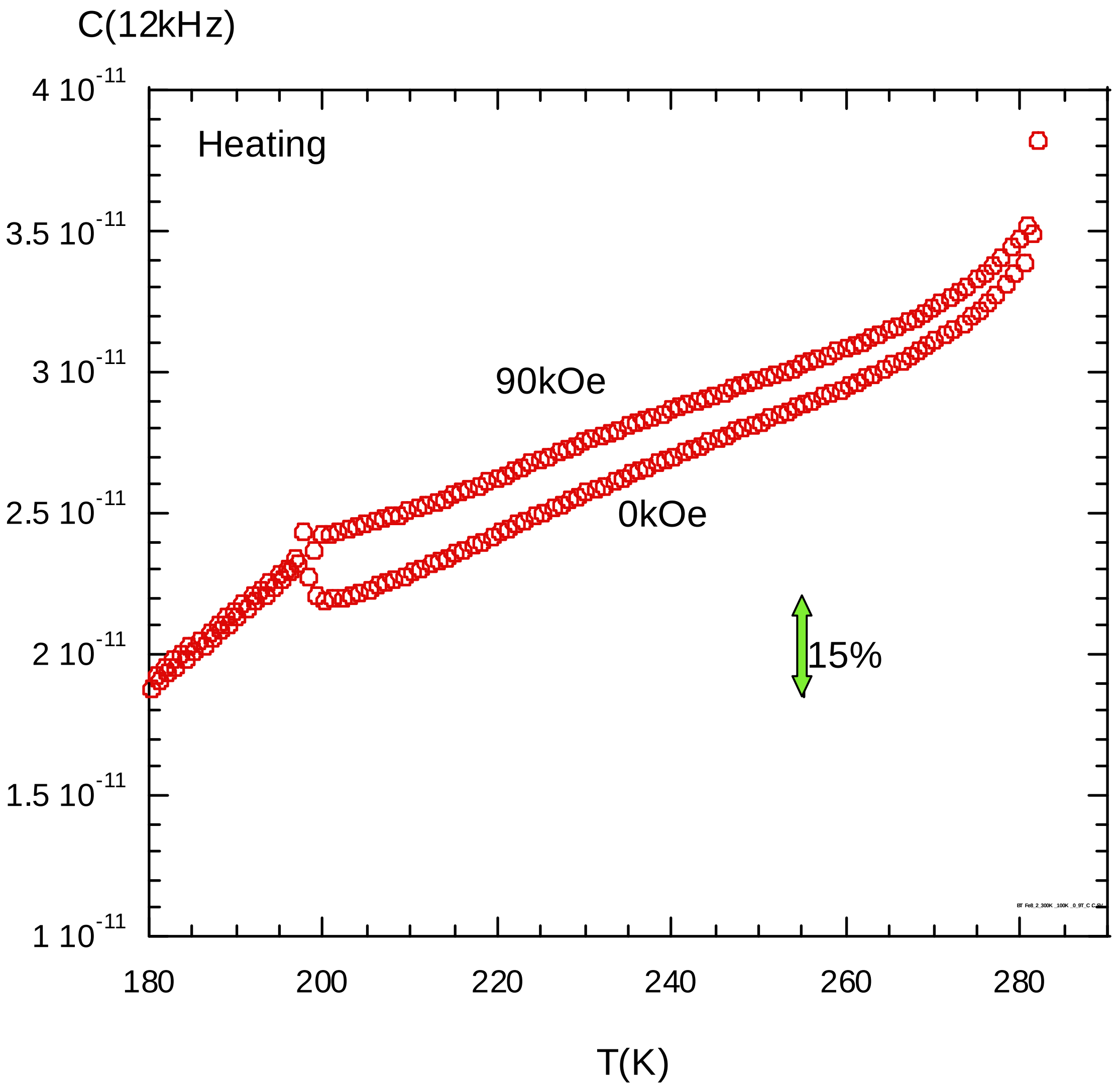

Figure 3c

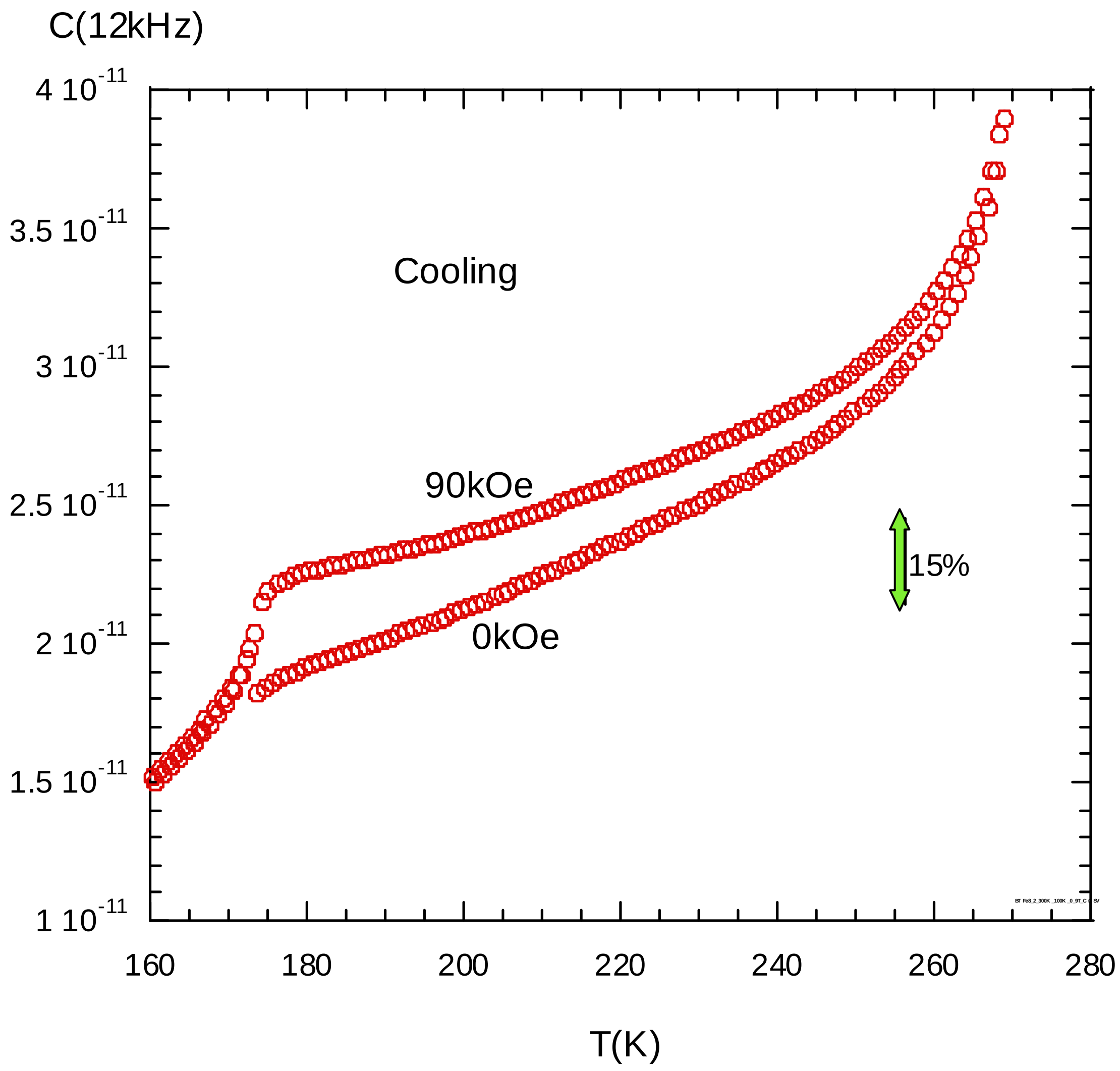

Figure 3d